\documentclass[12pt]{article}
\usepackage{graphicx} 
\usepackage{amsmath}
\usepackage{amssymb}
\date {}
\begin{document}
\title{Size effect on thermodynamic properties of free nanocrystals }
\author{A.\ I. Karasevskii\footnote{ \textit{E-mail address:} akaras@imp.kiev.ua} \ and
V.\ V. Lubashenko\footnote{ \textit{E-mail address:}
  vilu@imp.kiev.ua} \\G.\ V. Kurdyumov Institute for Metal Physics
\\ 36 Vernadsky str., Kiev, 03142, Ukraine}

\date{ }

\maketitle

\begin{abstract}We demonstrate that the discrete character of the vibrational spectrum of a small crystal
accounts for size dependence of its thermodynamic properties and
melting temperature. Using a  self-consistent statistical method
[Phys.~Rev.~B \textbf{ 66}, 054302 (2002)]  we derive the Gibbs
free energy of free nanocrystalline plates and calculate the
thermodynamic parameters as functions of plate thickness for Cu.
\end{abstract}


\section{\label{sec:1}Introduction}

The effect of  depression of melting point was first predicted
for free nanoparticles by Pawlow  \cite{Pawlow} from an analysis
of their thermodynamic properties. Further, it was proved
experimentally in transmission electron microscopy studies of
granular films of Pb, Sn, and Bi \cite{Takagi}. Later, a
relationship between the melting temperature of small metallic
species and their size was established on the basis of results of
electron microscopy studies of   Sn  \cite{Wronski}, Pb and In
\cite{Coombes}, Ag, Cu, Al and Ge \cite{Gladkich} thin films. A
comprehensive analysis of electron diffraction data on melting of
Au nanoparticles  and their comparison with predictions of
phenomenological approaches were carried out in
Ref.~\cite{Buffat}.

The electron diffraction methods, providing only the visual
picture of melting of nanocrystals,  allow one to relate the
melting temperature to the particle size, but they give no
information about  energetic properties of a nanosystem. In order
to get more insight into nature of melting of nanocrystals, these
methods are often combined with various nanocalorimetric
techniques, thus providing a possibility of estimation of   both
the melting temperature and the latent heat of fusion of
nanocrystals \cite{Willens}--\cite{Breaux}.

Early theoretical studies of thermodynamics of nanocrystals were
focused on attempts to explain the  strong size dependence of the
melting temperature. In thermodynamic consideration of an
equilibrium between solid and liquid  phases, an important role
was ascribed to the surface of the nanoparticle
\cite{Pawlow,Coombes,Buffat,Hanszen,Couchman}.  Some
phenomenological models of melting of free nanoparticles employed
the effect of surface melting inherent in the most of solids. For
example, the liquid-layer model \cite{Wronski,Peters} postulates
formation of a quasi-liquid layer at the surface of a nanocrystal
at temperature below its melting point. Among recent theoretical
approaches to description of size-dependent melting of
nanocrystals, we would  like to mention the liquid-drop model
\cite{Nanda} taking into account reduction of the average
cohesion energy of atoms of a nanoparticle with reduction of its
size and using an earlier proposed relation between the melting
temperature and the cohesion energy \cite{Tateno}.

Recently it was reported that thermodynamic properties of
nanocrystals, such as cohesion energy \cite{Yang2007,Qi2002},
Debye temperature \cite{Yang2007,Kastle,Yang}, activation energy
of diffusion \cite{Shibita, Jiang}, vacancy formation energy
\cite{Korhonen,Zhang2005} etc. display also  size dependence.
These experimental facts point to size influence  on statistical
characteristics of  atoms.

A considerable advance in computational capacity made for the
last decade has enabled  first-principles simulation of dynamical
behavior and energetic characteristics of systems consisting of
more than $10^5$ atoms \cite{Wang}--\cite{Baletto}, i.e.
nanocrystals in the so-called mesoscale regime.

A comprehensive molecular dynamics studies of thermodynamics  of
nanosized systems was carried out in Ref. \cite{Delogu2005} for
spherical Cu nanocrystals. The obtained  results demonstrate that
both melting temperature and the latent heat of fusion  decrease
smoothly as particle radius decreases. The analysis of energetic
and structural properties of nanocrystals revealed existence of a
rather thin surface layer where average potential energy and
root-mean-square (rms) displacements of atoms are largely
different from that of the core region. The rms displacements of
atoms in the surface layer increase dramatically at temperature
below the bulk melting point of the nanoparticle $T_m$,
indicating melting of this layer. The thickness of the surface
layer is relatively small and almost independent of temperature.
The rms displacements of bulklike atoms of the nanocrystal rise
linearly with temperature, and melting of the inner region of the
particle occurs when the Lindemann criterion is fulfilled.
Therefore, the molecular dynamics analysis of thermodynamic
properties of nanocrystals \cite{Delogu2005,Delogu2007} revealed
that a nanosized particle should be regarded as a structurally
and energetically heterogeneous system.

According to results obtained in Ref. \cite{Delogu2005}, the
behavior of the average potential  energy of atoms of a
nanocrystal exhibits essentially nonlinear rise near $T_m$. As
shown in Refs. \cite{KL2005,KHL2005} for the bulk solids, such a
behavior is attributed to evolution of anharmonic instability of
the phonon system of the crystal which is directly related to the
melting transition.

In this work we study size-dependent modification of
the phonon spectrum   of crystals and its influence on thermodynamic properties. For the sake of simplicity, we restrict ourselves with the case of a nanocrystalline plate. We show that the vibrational spectrum of such a system is substantially discrete and size-dependent. The phonon spectrum determines the parameters of statistical distribution of atomic displacements of the crystal. Thus, the average values of energetic properties of the crystal exhibit size dependence.
In our opinion, this
is an important mechanism responsible for depression of the melting
point and variation of thermodynamic properties of  free
nanocrystals with decreasing of their size. It is necessary to note that influence of
discreteness of the spectrum of eigenvalues on energetic
characteristics of many-particle finite systems was considered
previously for atomic nuclei \cite{Gurvits} and nanoparticles of
degenerated semiconductors \cite{Krivoglaz}. In these studies it
was demonstrated that corrections due to the discreteness of the
spectrum are proportional to the inverse size.

\section{\label{sec:2}Free energy of a crystal}

To examine size influence on thermodynamics of nanocrystals,  we
use a self-consistent statistical  method \cite{KL2002}
developed  to compute thermodynamic properties of simple  solids.
The method consists, first, in derivation of a binary
distribution function of atomic coordinates and, second, in a
variational procedure of computation of interatomic distance and
effective parameters of quasi-elastic bonds between atoms of the
crystal. It was shown that the phonon spectrum of the crystal
determines parameters of the  distribution function and,
therefore, the average values of energetic characteristics of the
crystal. Therefore, this approach allows one to clarify how
size-dependent modification of the phonon spectrum of a small
crystal affects its thermodynamic properties. The method has been
applied to description of thermal characteristics of the rare gas
crystals (RGC) \cite{KL2002}--\cite{KH2003b} and some simple fcc
metals \cite{KL2004}. In the framework of this method we also
computed the formation energy of a vacancy in the RGC as a
function of temperature and demonstrated that approaching the
melting point (which is assumed to be directly related to  the
point of anharmonic instability) is accompanied by dramatic
reduction of the energy required to create structure defects of
the lattice \cite{KL2005,KHL2005,KL2007}. As in our previous
studies, we assume the interatomic interaction to be pairwise and
approximated  by the Morse potential
\begin{equation}
u(r)=A \left[ e^{-2 \alpha (r-R_0)}-2 e^{-\alpha (r-R_0)}\right].
\label{Morse}
\end{equation}
The parameters $A$, $R_0$, and $\alpha$ have been determined so
that calculated values of interatomic distance, bulk modulus, and
sublimation energy at zero temperature fitted the corresponding
experimental data. These parameters  are given in
Ref.~\cite{KL2005} for the RGC and in Ref.~\cite{KL2004} for Cu
and Ag.

In this work we  restrict ourselves with the high-temperature limit when the distribution function of atomic displacements in a simple crystal may be represented as a product of one-particle Gaussian functions \cite{KL2005} given by
\begin{equation}
f(\mathbf{q}_i)=C \exp \left( -\frac{\alpha^2 c^2 q_i^2 n(\tau)}{\tau }
\right), \label{f}
\end{equation}
where $\mathbf{q}_i$ is the displacement of an atom from the $i$th lattice
site, $\tau=k_B T/A$ is the reduced temperature,  $c$ is a
dimensionless effective parameter of quasi-elastic bond of
neighboring atoms, and the coefficient $n(\tau)$ determines contribution of the phonon spectrum to the distribution width. At high temperature it is expanded into a power series as
\begin{equation}
n(\tau)=\sum_{l=0}^{\infty}(-1)^l n_l \left(\frac{c \Lambda}{\tau}\right)^{2l},
\label{n(tau)}\end{equation}
where
\begin{equation}
 \Lambda=\frac{\hbar \alpha}{\sqrt{MA}}
\end{equation}
is the de Boer parameter for the Morse potential, $M$ is atomic mass. The condition for the applicability of the high-temperature approximation is given by inequality  $\tau > c \Lambda$.
Coefficients $n_l$ are determined by the phonon spectrum of the crystal \cite{KL2002}. For example,
\begin{align}
n_0 &=\frac1{2} \sum_{j} \int  \tilde{\omega}_{j}^2(\mathbf{K}) e_{j x}^2(\mathbf{K})\, d \mathbf{K},\label{n_0}
\\ n_1 &=\frac{1}{24}\sum_j \int
\tilde{\omega}_{j}^4(\mathbf{K})  e_{j x}^2 \, d\mathbf{K},\label{n_1}
\end{align}
where $\mathbf{K}$ is reduced wave vector \cite{KL2002}, \begin{equation}
\widetilde{\omega}_j(\mathbf{k})=\left(\frac{M}{A \alpha^2 c^2}\right)^{1/2}
\,\omega_{j} (\mathbf{k})
\end{equation}
is the reduced frequency  of a phonon  with wave vector
$\mathbf{k}$ and branch $j$, and  $e_{j x}(\mathbf{k})$ is  the
projection of  a phonon polarization vector  on the $x$ axis. The
integration in Eqs.~(\ref{n_0}) and (\ref{n_1}) is carried out
over the first Brillouin zone. For a bulk fcc crystal, $n_0 =2$,
$n_1 =5/6$. Eqs. (\ref{n_0})--(\ref{n_1}) were derived in
Ref.~\cite{KL2002} for the bulk crystals, but, according to
Ref.~\cite{Feynmann}, such an expression is applicable to any
finite system of bound harmonic oscillators.

In the high-temperature approximation the Gibbs free energy
$\Phi$ of  a simple fcc crystal consisting  of $N$ atoms  can be
written as  \cite{KL2002,KL2004,KL2005}
\begin{multline}
g(\tau,p,c,b)=\frac{\Phi}{AN}=\left(\frac{\tau}{3}+3 \tau \log
\frac{c \Lambda}{\tau} \right)+\frac{z}{2} \left[e^{-2 b+
\tfrac{2\tau}{n(\tau) c^2} }- 2 e^{-b+\tfrac{\tau}{2 n(\tau) c^2}}
\right]\\-\frac{a_3 \tau^2}{c^6}\left[e^{-2 b+
\tfrac{2\tau}{n(\tau) c^2} }- \frac14
\,e^{-b+\tfrac{\tau}{2n(\tau) c^2}} \right]^2-\varkappa_l
\frac{\epsilon}{A}\left( \frac{\sigma}{R} \right)^6+p\alpha^3 v
.\label{g}\end{multline}
Here $R$ is the nearest neighbor
distance,  $b=\alpha (R-R_0)$ is a reduced lattice expansion,
$z=12$ is the coordination number, $p=P/\alpha^3 A$ is reduced
external pressure, and  $v=R^3/\sqrt{2}$ is volume per atom.

The first term in (\ref{g}) determines the entropy part of the
free energy of atomic vibrations in the crystal, the second
term represents the average potential energy of interaction of
 neighboring atoms. The third term  determines a contribution to the free
energy of the crystal due to the cubic anharmonicity of collective
atomic vibrations, evaluated in the second order of the
perturbation theory ($a_3 \approx 2.31$ for the RGC
\cite{KL2005,KHL2005},  $a_3 \approx 2.9$ for Cu, and $a_3
\approx 2.0$ for Ag \cite{KL2004}). The next term  takes into
account the long-range attraction between atoms of a molecular
crystal, where $\varepsilon$ and $\sigma$ are the parameters of
the Lennard-Jones potential and $\varkappa_l=4.91$ for the fcc
lattice \cite{KL2005}. In the case of simple metals, this term
should be replaced with an electron gas contribution to the
kinetic energy of the crystal \cite{KL2004},
\begin{equation}
\varepsilon_{\mathrm{el}}=\frac35\, \left(\frac{3 \pi^2 }{2}
 \right)^{2/3} \frac{\hbar^2}{m R^2 A}, \label{F_el}
\end{equation}
where $m$ is the effective electron mass.

\section{\label{sec:3}Vibrational spectrum of a thin plate}

To determine size contribution to the free energy of a
nanocrystal of size $h$ ($h \gg R$), we  proceed from the
assumption that the size-dependent modification of the
vibrational spectrum is most pronounced in its long-wave part,
with the wave vectors $k \sim \pi/h$ or $kR \le 1$. Such a case
corresponds to the elastic vibrations of the medium and is
described by wave equations of theory of elasticity. The
consideration of the size dependence of thermodynamic properties
of nanocrystals will be carried out for the simplest type of
nanocrystals,  a thin plate. In this case, the dispersion
relations have a rather simple appearance. A consistent
consideration of elastic vibrations and propagation of elastic
waves in plates was originally performed by Lamb \cite{Lamb} in
the framework of general theory of elasticity.

Let us consider a plate of thickness $h$, with free surfaces
parallel to the $(x,y)$ plane  and the origin taken at the center
of the plate. The $z$ axis is normal to the free surfaces $z=\pm
h/2$. The system is assumed to be of macroscopic size in the
$(x,y)$ plane, so that vibrations can propagate as plane waves in
the $x$ and $y$ directions. Note that we do not consider
vibrations of the plate as a whole. The components of the stress
tensor  vanish at the free surfaces,
\begin{equation}\begin{aligned}
\left.\sigma_{zx}\right|_{\pm h/2}&=0,\\
\left.\sigma_{zz}\right|_{\pm h/2}&=0,\\ \left.\sigma_{zy}\right|_{\pm h/2}&=0.
\end{aligned}\label{bound_cond}\end{equation}
For such a system, the displacement vector $\mathbf{u}$ can be
represented as superposition of waves of the horizontal
polarization, with $u_x=u_z=0$,  $u_y=u$, and waves of the
vertical polarization with $u_x\ne 0$, $u_z\ne 0$, and  $u_y=0$.
It is important that these two types of waves do not mix at the
boundaries, so that they can be considered separately (see, e.g.
Refs.~\cite{Lamb,Love}).

Waves of the horizontal polarization are described by ordinary wave equations for the displacement $u_y$,
\begin{equation}
\frac{\partial^2 u_y}{\partial x^2}+\frac{\partial^2 u_y}{\partial z^2}+\kappa^2 u_y=0, \label{wave_eq}\end{equation}
where $\kappa=\omega/c_t$, $\omega$ is an eigenfrequency, $c_t$ is transverse sound velocity. A  solution of (\ref{wave_eq}) in a particular case of a plane wave propagating along the $x$ direction is given by
\begin{equation}u_y=C \cos (\beta z-\delta)\, e^{i \xi x}.\label{u_y}\end{equation}
Parameters $\beta$ and $\delta$ are found from both the boundary
conditions (\ref{bound_cond})  and the symmetry reasons allowing
two types of the solutions, a symmetrical one,
$u_y(z=h/2)=u_y(z=-h/2)$, and an antisymmetrical one,
$u_y(z=h/2)=-u_y(z=-h/2)$. These solutions lead generally to a
dispersion relation given by
\begin{equation}\omega_{t}=c_t k_n,\label{disp_rel_horizont-1}\end{equation}
with
\begin{equation}k_n=\sqrt{\xi^2+\zeta^2+\left(\frac{n\pi}{h}\right)^2},
\quad n=1,2,3,\dots \label{disp_rel_horizont}\end{equation} Here
$\xi$ and $\zeta$ are, respectively, projections of the wave
vector  on the $x$ and $y$ axes  for plane waves propagating in
the $xy$ plane. The projection of  the wave vector  on the $z$
axis takes on discrete values,
\[k_z=\frac{n\pi}{h}.\]

In the case of waves of vertical polarization, the displacement
is expressed in terms of  scalar $\varphi$ and vector
$\mathbf{\Psi}$ potentials,
\begin{equation}\mathbf{u}=\mathrm{grad}\, \varphi+\mathrm{rot}\, \mathbf{\Psi} \label{u_vert}.\end{equation}
The potential $\mathbf{\Psi}$ can be chosen so that $\Psi_y=\Psi$
and $\Psi_x=\Psi_z=0$. Then  $\varphi$ and $\Psi$ satisfy the
wave equations,
\begin{align} \Delta \varphi+k^2 \varphi&=0, \label{wave_eq_phi}\\
\Delta \Psi+\kappa^2 \Psi&=0, \label{wave_eq_psi}\end{align}
where $k=\omega/c_l$, $c_l$ is longitudinal sound velocity.
The solutions of Eqs.~(\ref{wave_eq_phi})--(\ref{wave_eq_psi}) are given by
\begin{align} \varphi&=C \cos (\alpha z-\delta_l)\, e^{i \xi x},\label{phi}\\
\Psi&=D \cos (\beta z-\delta_t)\, e^{i \xi x},\label{psi} \end{align}
where $\alpha$ and $\beta$ are real values given by
\begin{align}\alpha&= \sqrt{k^2-\xi^2},\label{alpha}\\
\beta &= \sqrt{\kappa^2-\xi^2}.\label{beta} \end{align}
 In terms of these notations, the boundary conditions (\ref{bound_cond}) are rewritten as
\begin{align}
\left. \frac{\partial \Psi}{\partial z}-i p \,\varphi \right|_{z=\pm h/2}&=0, \label{bound_cond-1}\\
\left. \frac{\partial \varphi}{\partial z}+i p \, \Psi \right|_{z=\pm h/2}&=0, \label{bound_cond-2}
\end{align}
where $p=(\xi^2-\beta^2)/(2\xi)$. Hence we get the general dispersion relation for the acoustic part of the vibrational spectrum of the nanocrystal,
\begin{equation} \tan \left(\frac{\alpha h}{2}-\delta_l^0 \right)\, \tan \left(\frac{\beta h}{2}-\delta_t^0 \right)=\frac{p^2}{\alpha \beta}.\label{disp_rel}\end{equation}
From the symmetry reasons, the projections of the displacement vectors $u_x$ and $u_z$ should be either symmetrical or antisymmetrical. At the symmetrical conditions
\[\delta_l^0 =m \pi, \quad \delta_t^0 =(2n+1)\pi/2, \quad  m,n=0,1,2, \dots \]
the dispersion relation (\ref{disp_rel}) appears as
\begin{equation}\cot \frac{\alpha h}{2}\, \tan \frac{\beta h}{2} =-\frac{p^2}{\alpha \beta}. \label{disp_rel-sym}\end{equation}
The antisymmetrical condition is specified by
\[\delta_l^0 =(2n+1) \pi/2, \quad \delta_t^0 =m\pi, \quad m,n=0,1,2, \dots \]
leading to
\begin{equation}\tan \frac{\alpha h}{2}\,  \cot\frac{\beta h}{2} =-\frac{p^2}{\alpha \beta}. \label{disp_rel-antisym}\end{equation}
Note that satisfying the symmetrical condition for $u_x$ implies satisfying the antisymmetrical condition for $u_z$, and viceversa.

\begin{figure}[ht]
\includegraphics[width=12cm]{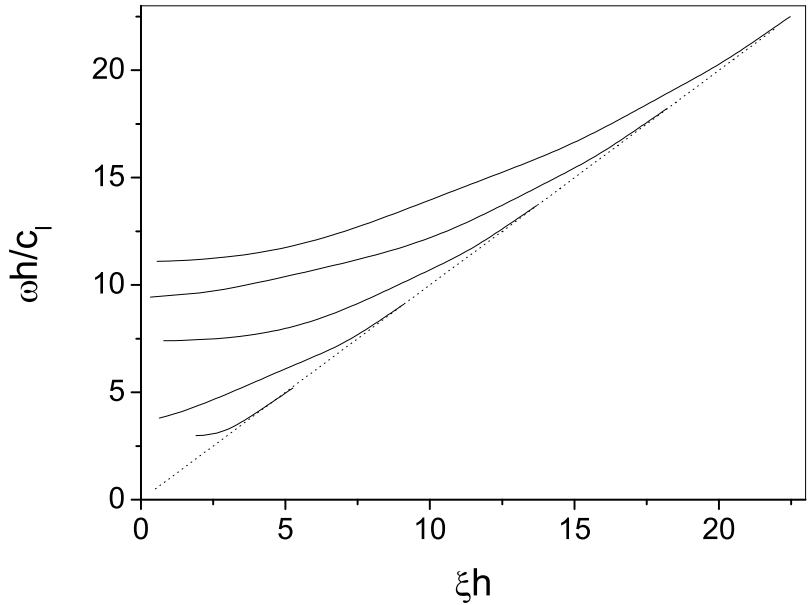}
\caption{\label{fig:1}Dispersion curves of longitudinal vibrations of a  thin plate for $n=1,\dots 5$. The dash line corresponds to the dispersion relation $\omega=c_l \xi$ for the bulk crystal.
}
\end{figure}

In Fig. \ref{fig:1} we plotted dispersion curves for the
longitudinal characteristic  vibrations of a thin plate obtained
from (\ref{disp_rel-sym}). At a rather small thickness $h$ of the
plate, the long-wave part ($kR\le 1$) of the vibrational spectrum
of the nanocrystal differs substantially from that of the bulk
crystal, described by the linear relation between the frequency
and the wave vector, $\omega_j=c_j \xi$ ($j=l,t$). The spectrum
of the thin plate  splits to separate vibration branches. For
each  branch, there is an integer number of wavelengths going
across the plate thickness. The dispersion curves are well
approximated by parabolas and can be analytically represented as
\begin{equation}\omega_n^2(\xi,\zeta)=\omega_{n,b}^2+\alpha_n c_l^2 (\xi^2+\zeta^2). \label{omega_n}\end{equation}
Here $\omega_{n,b}$ is the minimal value of the frequency of the
$n$th vibration branch, $\alpha_n\approx 1$.  For each of  both
symmetrical (\ref{disp_rel-sym})  and antisymmetrical
(\ref{disp_rel-antisym}) cases we have two types of solutions for
$\omega_{n,b}$ corresponding to longitudinal and transverse
standing waves with the frequencies generally written as
\begin{equation} \omega_{n,b}^j \approx \frac{\pi n}{h}\, c_j, \quad j=l,t. \label{omega_nb}\end{equation}
As in the case of waves of the horizontal  polarization
(\ref{disp_rel_horizont}), the wave vector for the both
longitudinal and transverse  waves of the vertical polarizations
is given by Eq.~(\ref{disp_rel_horizont}), too.

A similarity of the expression for $(\omega_{n,b}^j )^2$ to the
formula for the energy of a  quantum particle in a
one-dimensional potential well \cite{Landau} is explained by the
wave nature of both quantum particles and vibrational modes,
exhibiting discreteness of the eigenvalues in the case of a small
size of the system.

For the bulk fcc crystals, the  first Brillouin zone has the form
of a truncated octahedron and may be well approximated by a
sphere of radius $k_s^0=(6\pi^2/v)^{1/3}$. The radius $k_s^0$ is
determined by the requirement that the number of independent
values of the wave vector falling within the sphere is equal to
the number of atoms of the simple lattice.

In the case of nanocrystals, evaluation of the radius $k_s$ of the spherical Brillouin zone requires taking into account the discrete character of $k_z$ in replacing summation over $\mathbf{k}$ by integration. This can be done using the Euler-Maclaurin formula \cite{Madelung}
\begin{equation}\sum_{k_{min}}^{k_{max}} f(k)=\int_{k_{min}}^{k_{max}} f(t) \, dt+\frac12 \left[f(k_{min})+f(k_{max}) \right] +\dots .\label{Euler-Maclaurin}\end{equation}
Then we get
\begin{equation}k_s=k_s^0 \left(1+\frac{\pi}{4(k_s^0 R)} \frac{R}{h} \right). \label{k_B}\end{equation}
Presence of a size-dependent factor in (\ref{k_B}) results in
appearance of  similar corrections  to the coefficients $n_l$ in
(\ref{n(tau)}),
\begin{equation} n_l (h)= n_l^0 \left(1+\gamma_l \, \frac{R}{h} \right),\end{equation}
where $n_l^0$ is the bulk value of $n_l$, the first two
coefficients $\gamma_l$ are $\gamma_0=0.68$ and $\gamma_1=0.74$.
As  will be shown below, the size dependence of the coefficients
$n_l$ leads to a number of physical effects that should be observed in
nanocrystals.

It should be noted that the above procedure of calculation of
the coefficients $\gamma_l$ seems to give overestimated values,
because Eq.~(\ref{k_B}) for the quasi-discrete wave vector is
valid, strictly speaking, only in the range $kR \le 1$, sensible
to the particle size. Near the boundaries of the Brillouin zone,
the values of $\mathbf{k}$ are determined by the structure of the
crystal lattice and are slightly  dependent of the crystal size.
However, the structure of Eq.~(\ref{k_B}) should remain, on the
whole, the same, so that $\gamma_l$ may be considered as
parameters of the theory.

\section{\label{sec:4}Thermodynamic properties of nanocrystals}

The equilibrium values of  the quasi-elastic bond parameter $c_0$
and reduced lattice expansion $b_0$ are  determined from
minimization of the Gibbs function (\ref{g}) with respect to $c$
and $b$,
\begin{equation}\left.\frac{\partial g}{\partial c}\right|_{\tau,p,b}=0, \quad
\left.\frac{\partial g}{\partial
b}\right|_{\tau,p,c}=0.\label{equilib}\end{equation} The
condition for the equilibrium values of the variational
parameters to exist is that
\begin{equation}D=\det \left|
\begin{array}{ll} g_{cc}''& g_{cb}'' \\ g_{bc}'' & g_{bb}'' \end{array}
 \right|>0.\label{det}\end{equation}
The sufficient condition $D=0$  defines the point of anharmonic
instability of the crystal where the minimum of $g(\tau,p,c,b)$
with respect to $c$ and $b$ disappears.

\begin{figure}[ht]
\includegraphics[width=12cm]{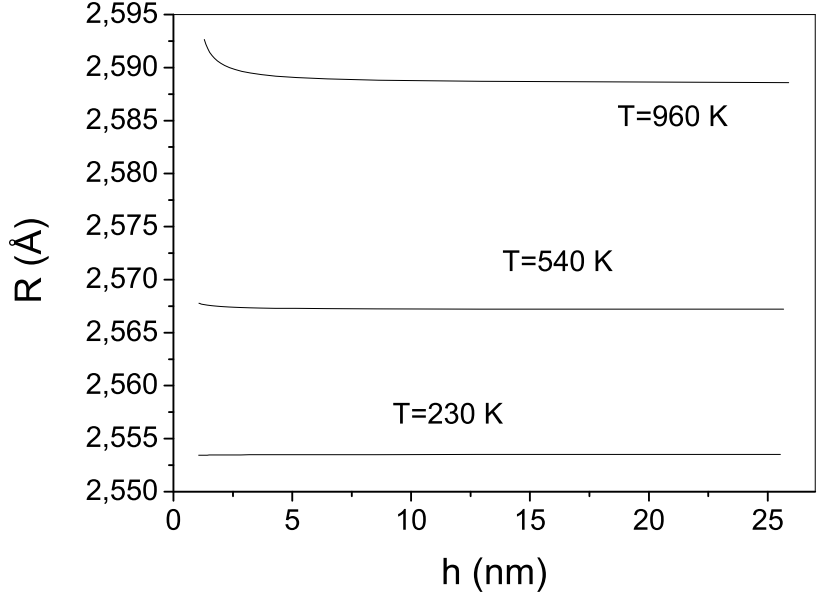}
\caption{\label{fig:2}Size dependence of the interatomic distance
in Cu nanocrystals  at different temperatures. }
\end{figure}

In the case of nanocrystals, the Gibbs free energy depends,
besides temperature $\tau$ and pressure $p$, on  the crystal size
$h$. As a result, a number of thermodynamic characteristics of
the crystal display size dependence. For example, the Debye
temperature is directly proportional to the equilibrium value of
the parameter $c_0(\tau,p,h)$ and is written as
\cite{KH2003a,KH2003b}
\begin{equation}\Theta_D \approx (6 \sqrt2 \pi^2)^{1/3} \langle \kappa_{j \mathbf{k}} \rangle  A \Lambda c_0(\tau,p,h) /k_B,  \label{Debye}\end{equation}
where coefficients $ \kappa_{j \mathbf{k}} =\omega_{j \mathbf{k}}/(kR)$ account for different polarizations of the acoustic waves, and  $\langle \kappa_{j \mathbf{k}} \rangle \approx 0.67$ for the fcc crystals.   The volume and temperature dependence of $\Theta_D$ determines the Gr{\"u}neisen parameter,
\begin{equation}
\gamma_{\mathrm{G}}=-\frac{ (\partial \ln \Theta_D/\partial
\ln V )_T}{ 1- (\partial \ln \Theta_D/\partial
\ln T)_V}. \label{Grun}
\end{equation}
The  equilibrium value of  $b_0(\tau,p,h)$ governs the
interatomic distance, the coefficient of thermal expansion, and
the bulk modulus of the nanocrystal. All these properties would
depend on the crystal size.

If the melting temperature $T_m$ is assumed to  be identical to
the temperature of anharmonic instability of the crystal
\cite{KL2005,KHL2005}, then it can be shown in the framework of
the present statistical model  that $T_m$ is related to the
factor $n(\tau)$ as
\begin{equation}\tau_m=\frac{k_B T_m}{A} \sim \frac{1}{a_3 n^3(\tau).} \label{tau_m}\end{equation}
Hence it follows that even a slight increasing of $n(\tau)$ accompanying reduction of the crystal size results in appreciable depression of the melting temperature of the nanocrystal. Since $R/h \ll 1$, it follows from (\ref{tau_m}) that
\begin{equation}\frac{T_{m}}{T_m^b} =1-\frac{\beta}{3h},\label{T_M}
\end{equation}
where $T_m^b$ is the melting temperature of the bulk material,
and $\beta=9 \gamma_0 R$. This is the well-known relationship
between the relative melting temperature of a plate nanocrystal
and its inverse size (see, e.g.,   \cite{Nanda}). The calculated
value $\beta=1.59$~nm for Cu is somewhat higher than that
extracted from experimental data \cite{Nanda} ($\beta \approx 0.9
- 1.2$~nm).

In Fig.~\ref{fig:2}  we plotted the  interatomic distance in Cu nanocrystals versus particle size  calculated from (\ref{equilib}) at three values of temperature. Hereafter all theoretical results are presented for zero external pressure. A nanocrystal of size $h \approx 1.3$~nm melts at $T=960$~K, which explains some nonlinear rise of $R(h)$ in the top curve. These results comply with the assumption \cite{Nanda2001,Qi2005}  that the size-induced lattice contraction observed in spherical nanoparticles is due to capillary pressure only.

\begin{figure}[ht]
\includegraphics[width=12cm]{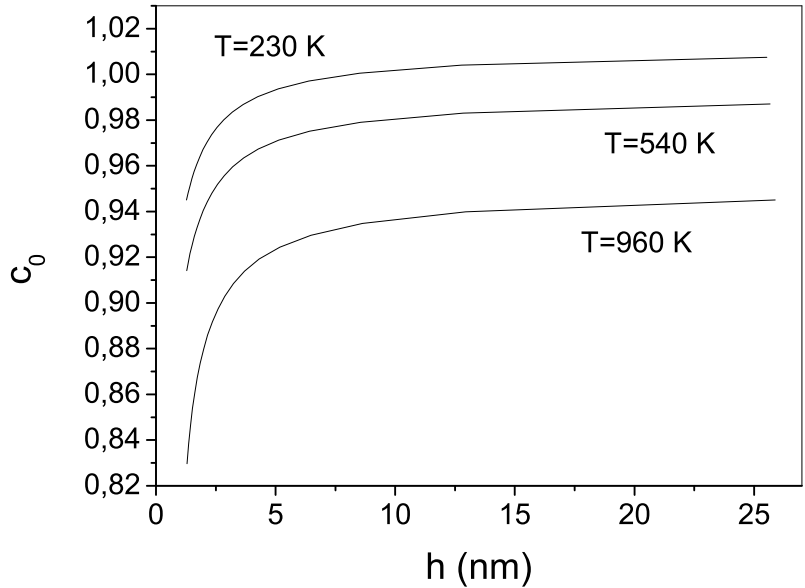}
\caption{\label{fig:3}Size dependence of the effective
quasi-elastic bond parameter $c_0(\tau,h)$  of Cu nanocrystals
at different temperatures. }
\end{figure}

While the interatomic distances in the considered nanocrystals
show no marked size dependence, the effective parameter
$c_0(\tau,p,h)$ of the quasi-elastic bond varies substantially
with $h$. Such dependence calculated for Cu nanocrystals  is
illustrated with Fig.~\ref{fig:3} which demonstrates that $c_0$
decreases nonlinearly as $h$ decreases even at $T \ll T_m (h)$.

\begin{figure}[ht]
\includegraphics[width=12cm]{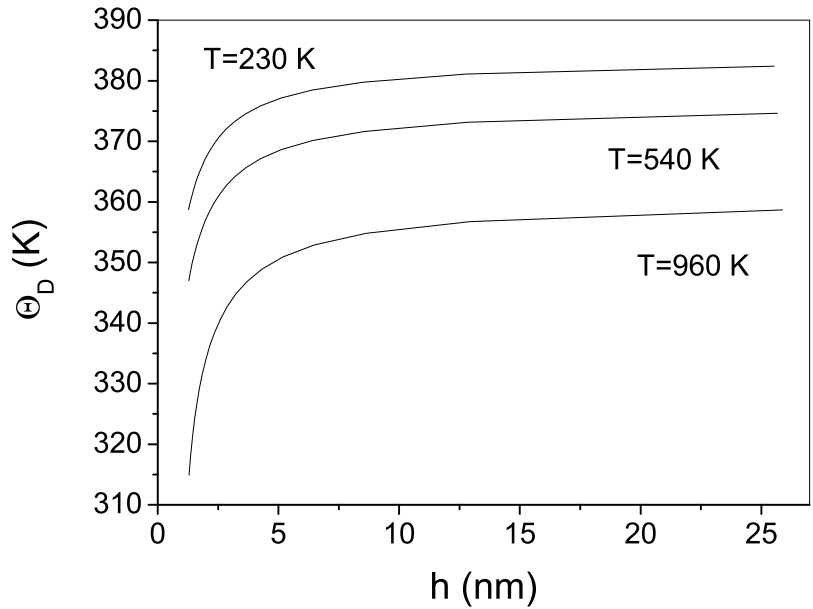}
\caption{\label{fig:4}Size dependence of the Debye temperature  of Cu nanocrystals  at different temperatures.
}
\end{figure}

Size dependence of the Debye temperature $\Theta_D$ of small
crystals was observed in a number of experimental studies
\cite{Yang2007,Kastle,Yang}. In the framework of the present
approach, $\Theta_D$ is proportional to the quasi-elastic bond
parameter $c_0$ (\ref{Debye}). The $\Theta_D(h)$ dependence
presented in Fig.~\ref{fig:4} agrees qualitatively with that
obtained experimentally, e.~g. for gold nanoparticles
\cite{Kastle}.

\begin{figure}[ht]
\includegraphics[width=12cm]{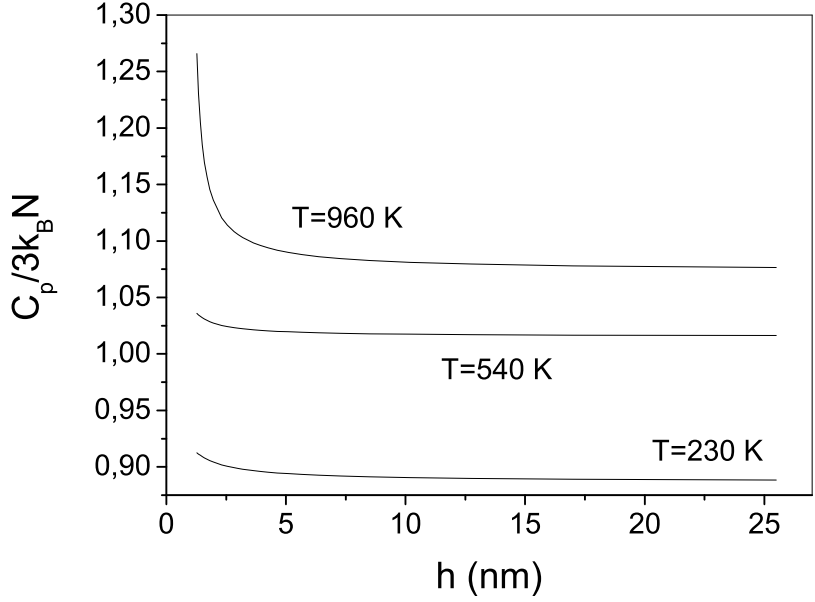}
\caption{\label{fig:5}Size dependence of the isobaric heat capacity  of Cu nanocrystals  at different temperatures.
}
\end{figure}

In many bulk solids,  the melting transition is preceded by anomalous behavior of temperature derivatives of thermodynamic functions (isobaric heat capacity, thermal expansion coefficient, Gr{\"u}neisen parameter) due to evolution of the anharmonic instability in the phonon subsystem of the crystal as $T \to T_m$ \cite{KL2005,KHL2005} (premelting phenomena). As seen from Fig.~\ref{fig:5} for the isobaric heat capacity $C_P$,  similar behavior should be also observed  when the melting point is approached via reduction of the crystal size at constant temperature (the top curve).

\begin{figure}[ht]
\includegraphics[width=12cm]{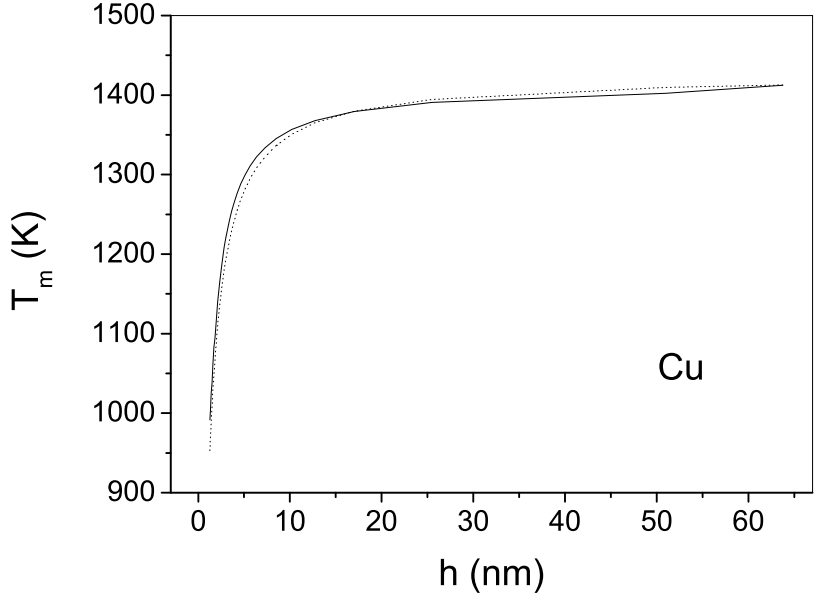}
\caption{\label{fig:6}Melting temperature of a free Cu thin plate   versus  its thickness calculated by minimization of the Gibbs free energy (\ref{g}) (solid curve) and from Eq.~(\ref{tau_m}) (dotted curve).
}
\end{figure}

In Fig.~\ref{fig:6}  we show the   melting temperature of a free Cu thin plate  as a function of  its thickness
calculated from Eqs.~(\ref{equilib}) and the condition $D=0$, compared with that obtained from Eq.~(\ref{tau_m}).

At present, the only generally recognized melting criterion is the empirical Lindemann rule suggesting that a solid melts when the rms displacement of atoms reaches a characteristic fraction of the interatomic distance. In terms of notations accepted in this work, the Lindemann ratio $\delta$ is given by
\begin{equation}
\delta^2=\frac{\langle q^2 \rangle}{R^2}=\frac{3 \tau}{4c_0^2(\tau) (\alpha
R)^2 g_t}. \label{lind}
\end{equation}
Here $g_t\approx 0.77$ is the  correlation smearing of the
distribution width of an atom at a lattice site \cite{KL2002}.

\begin{figure}[ht]
\includegraphics[width=12cm]{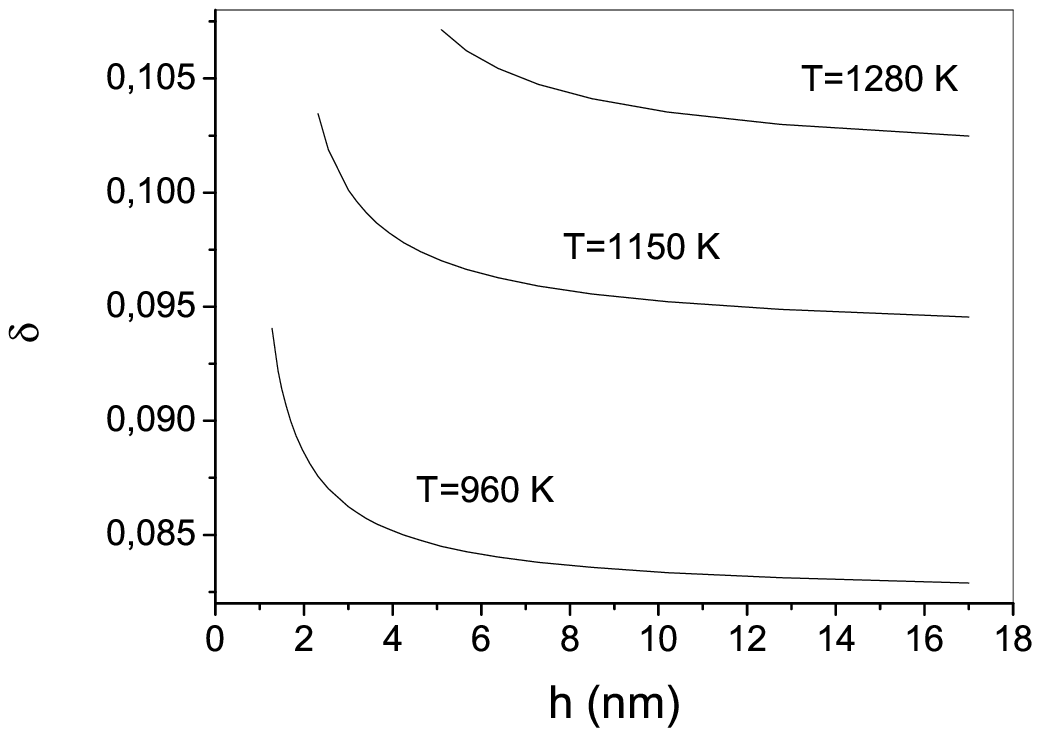}
\caption{\label{fig:7}The relative rms displacement  of atoms (the Lindemann ratio) of Cu thin plate versus its thickness  at different temperatures.
}
\end{figure}

The size dependence of the Lindemann ratio is presented in
Fig.~\ref{fig:7}. A dramatic increase of the rms displacement of atoms in the premelting region is caused by nonlinear
reduction of the quasi-elastic bond parameter $c_0(\tau,p,h)$
in the vicinity of the instability point (Fig.~\ref{fig:3}). The maximal value $\delta \sim 0.1$  depends only slightly on
$h$. Thus, as for the bulk solids, the premelting region is characterized by a nonlinear rise of some solid-state parameters of nanocrystals (isobaric heat capacity, the coefficient of thermal expansion, rms displacement etc.). It was also shown for the bulk case that the
formation energy of  the structural lattice defects exhibit a sharp drop just before $T_m$. All these premelting effects promote transition to a structurally disordered phase, i.e. melting.

\begin{figure}[ht]
\includegraphics[width=12cm]{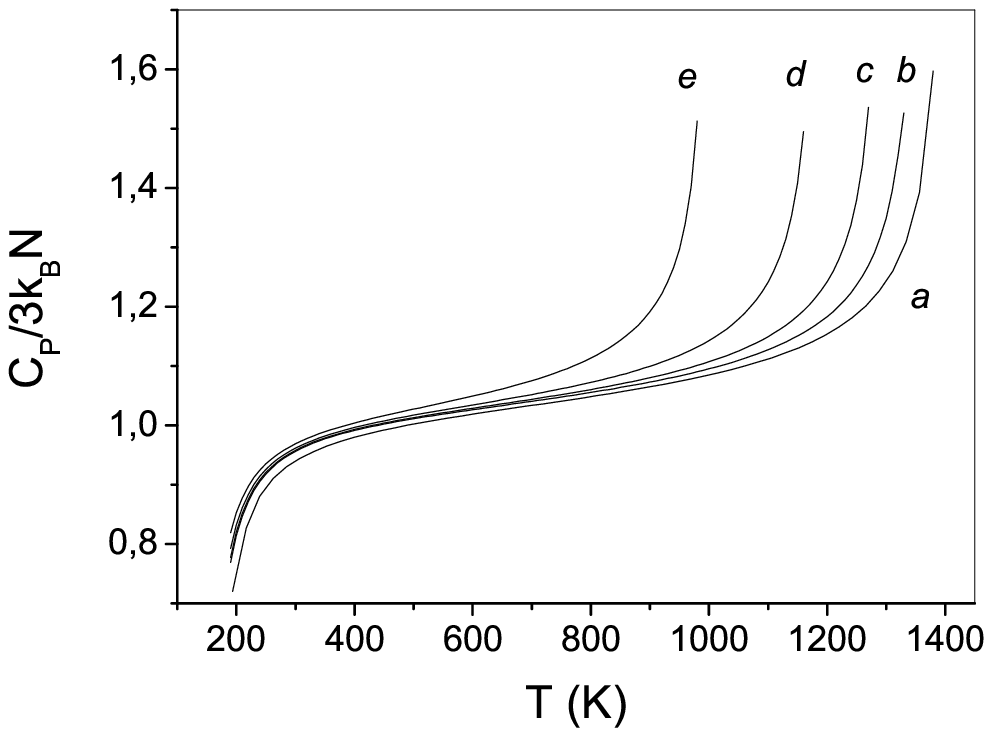}
\caption{\label{fig:8}The temperature dependence of the isobaric heat capacity of Cu nanocrystals: (a) bulk crystal; (b) $h=10.2$~nm; (c) $h=5.1$~nm;
(d) $h=2.55$~nm; (e) $h=1.26$~nm.}
\end{figure}

\begin{figure}[ht]
\includegraphics[width=12cm]{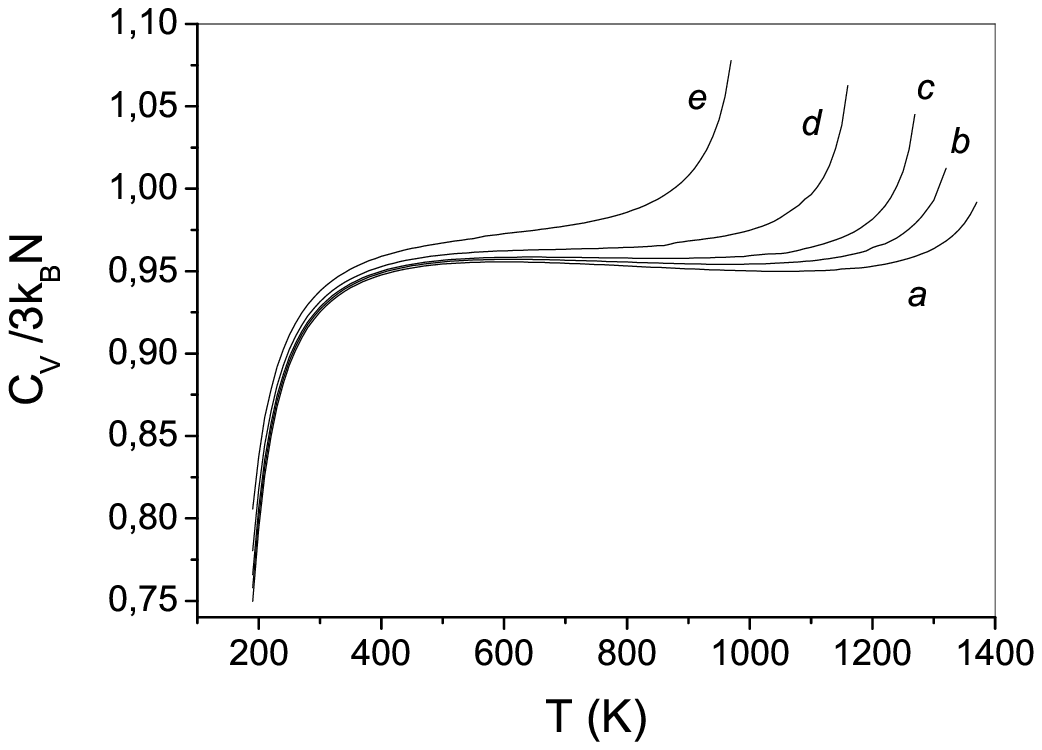}
\caption{\label{fig:9}The temperature dependence of the isochoric heat capacity of Cu nanocrystals: (a) bulk crystal; (b) $h=10.2$~nm; (c) $h=5.1$~nm;
(d) $h=2.55$~nm; (e) $h=1.26$~nm.}
\end{figure}

\begin{figure}[ht]
\includegraphics[width=12cm]{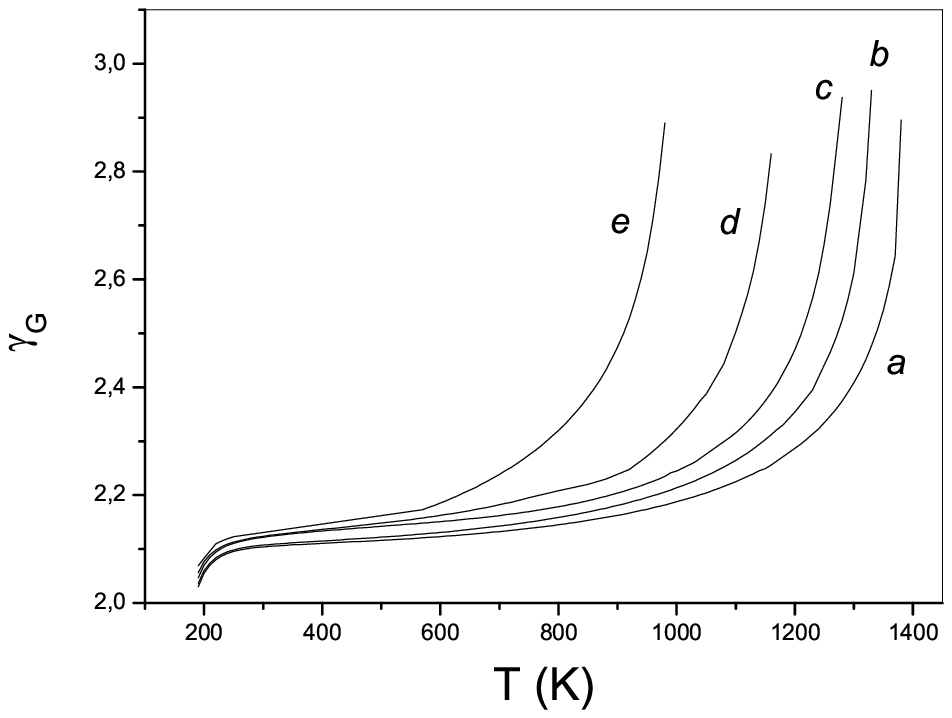}
\caption{\label{fig:10}The temperature dependence of the
Gr{\"u}neisen  parameter of Cu nanocrystals: (a) bulk crystal;
(b) $h=10.2$~nm; (c) $h=5.1$~nm; (d) $h=2.55$~nm; (e)
$h=1.26$~nm.}
\end{figure}

Fig.~\ref{fig:8}--\ref{fig:10} illustrate nonlinear behavior of thermodynamic properties of nanocrystals when the instability point is approached via increasing temperature at constant size. We see that the premelting range shifts as the crystal size decreases.

\begin{figure}[ht]
\includegraphics[width=12cm]{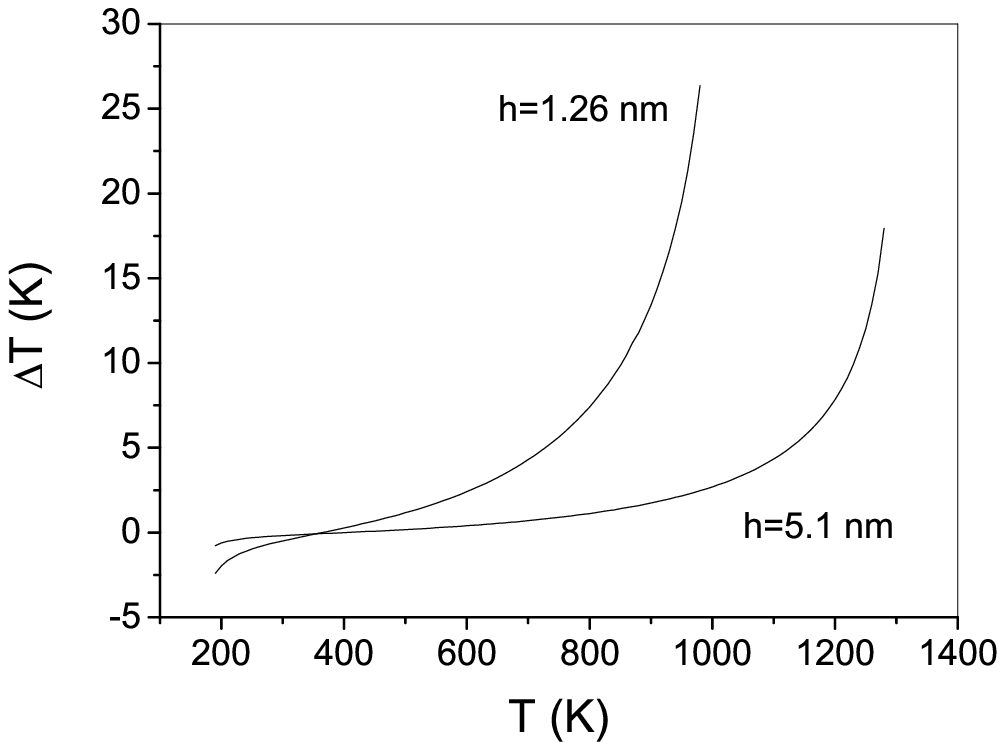}
\caption{\label{fig:11}A temperature change at mechanical contact of two identical Cu thin plates of thickness $h$  versus initial temperature.
}
\end{figure}

If two nanocrystals of the same thickness $h$ are brought into
tight mechanical  contact, the vibrational modes with wavelengths
larger than the characteristic thickness of the contact become
collectivized, and their spectrum becomes, roughly speaking, equal to the
spectrum of  a nanocrystal with
thickness $2h$ \cite{Physical_acoustics}. Breaking the contact,
i.~e. separating the two  plates, results in the reverse
rearrangement of their vibrational  spectra and is accompanied by
a change $\Delta \varepsilon_h$ in the energy per atom
$\varepsilon=g-\tau \partial g/\partial \tau$. Obviously,
$\Delta \varepsilon_h=-\Delta \varepsilon_{2h}$. If the
nanoparticles are isolated, the energy change would lead to a
change in the temperature of the nanocrystals, $\Delta T=\Delta
\varepsilon_{h}/c_V$, where $c_V$ is isochoric heat capacity per
atom of the nanocrystal. In Fig.~\ref{fig:11}   we plot a contact
temperature change   $\Delta T$ of two identical Cu thin plates
versus initial temperature calculated for two values of the plate
thickness, $h=5.1$~nm and $h=1.26$~nm.  As follows from the
calculations, the temperature change is maximal if the initial
system of separated nanoparticles is close to the melting point.

\section{\label{sec:5} Discussion}

So far there is a large amount of experimental data concerning
size influence on the bulk properties of nanoparticles, such as
cohesion energy \cite{Yang2007,Qi2002}, Debye temperature
\cite{Yang2007,Kastle,Yang}, and  activation energy of diffusion
\cite{Shibita,Jiang}. One of the most pronounced manifestations
of change of thermodynamic properties observed in nanosized
crystalline systems is the effect of reduction of the melting
temperature of free nanocrystals \cite{Takagi}--\cite{Breaux}. So
the principal problem of statistical description of
thermodynamics of nanocrystals is elucidation of the mechanism of
size effect on statistical characteristics of atoms in such
systems. Relying on the results of molecular dynamics simulations
of thermodynamic properties and melting of nanocrystals
\cite{Delogu2005,Delogu2007}, it is natural to conclude that a
thin (relative to the particle size) surface layer has only a
minor influence upon the bulk properties of the particles;
moreover, in the case of spherical particles an additional
capillary pressure should contribute to increasing of both the
melting temperature and the Debye temperature.

It is  shown in this work that an important size-dependent factor
governing thermodynamic behavior of a nanocrystal  is
discreteness of its vibrational spectrum. It leads to increasing
of the parameter $n(\tau)$ of the statistical distribution
function of atomic coordinates (\ref{f}) as crystal's size decreases. As a
consequence,   the average value of the interaction energy  of
atoms in the nanocrystal changes with its size. It is necessary to emphasize that such
influence on a crystalline system mediated by direct change of the
statistical distribution function is inherent only in
nanosystems. In this connection, it is worth noting a  specific character of thermodynamic response of a nanocrystal to variation
of its size. As the crystal's size decreases, the parameter
$c_0(\tau,p,h)$ of quasi-elastic bond of its atoms decreases
(Fig.~\ref{fig:3}), as well as the Debye temperature
(Fig.~\ref{fig:4}), while interatomic distance remains nearly
constant (Figs.~\ref{fig:2}). The melting temperature
(\ref{tau_m}) of the nanocrystal  is markedly decreased
(Figs.~\ref{fig:6}). Independently of the crystal size, the
melting transition occurs when the Lindemann criterion is
satisfied (Fig.~\ref{fig:7}). Along with a size-dependent shift
of the melting temperature, there is also a corresponding shift
of the premelting range  where thermodynamic
properties (isobaric heat capacity, coefficient of thermal
expansion etc.) display  nonlinear behavior.

\section*{Acknowledgement}
This  work was supported in part by Award no.~28/08-H  in the
framework of the Complex Program of Fundamental Investigations
``Nanosized systems, nanomaterials, nanotechnology'' of National
Academy of Sciences of Ukraine.

\end{document}